# An asymmetric shock wave in the 2006 outburst of the recurrent nova RS Ophiuchi

T. J. O'Brien[1], M. F. Bode[2], R. W. Porcas[3], T. W. B. Muxlow[1], S. P. S. Eyres[4], R. J. Beswick[1], S. T. Garrington[1], R. J. Davis[1] & A. Evans[5]

[1]Jodrell Bank Observatory, School of Physics & Astronomy, The University of Manchester, Macclesfield, Cheshire, SK11 9DL, UK. [2]Astrophysics Research Institute, Liverpool John Moores University, Twelve Quays House, Birkenhead, CH41 1LD, UK. [3]Max-Planck-Institut für Radioastronomie, Auf dem Hügel 69, D-53121 Bonn, Germany. [4]Centre for Astrophysics, University of Central Lancashire, Preston, PR1 2HE, UK. [5]Astrophysics Group, School of Physical & Geographical Sciences, Keele University, Staffordshire, ST5 5BG, UK.

**Nova outbursts[1] take place in binary star systems comprising a white dwarf and either a low-mass Sun-like star or, as in the case of the recurrent nova RS Ophiuchi[2], a red giant. Although the cause of these outbursts is known to be thermonuclear explosion of matter transferred from the companion onto the surface of the white dwarf[3], models of the previous (1985) outburst of RS Ophiuchi failed to adequately fit the X-ray evolution[4] and there was controversy over a single-epoch high-resolution radio image, which suggested that the remnant was bipolar[5,6] rather than spherical as modelled. Here we report the detection of spatially resolved structure in RS Ophiuchi from two weeks after its 12 February 2006 outburst. We track an expanding shock wave as it sweeps through the red giant wind, producing a remnant similar to that of a type II supernova but evolving over months rather than millennia[7]. As in supernova remnants, the radio emission is non-thermal (synchrotron emission), but asymmetries and multiple emission components clearly demonstrate that contrary to the assumptions of spherical symmetry in models of the 1985 explosion, the ejection is jet-like, collimated by the central binary whose orientation on the sky can be determined from these observations.**

During the previous outburst of RS Ophiuchi (RS Oph) in 1985 a campaign was organized incorporating observations ranging from radio to X-ray wavelengths. The results included the detection of bright, evolving X-ray emission from hot gas suggested to arise from the expanding shock wave[8]. This time we have monitored RS Oph from much earlier in the outburst, both in X-rays[9–13], and at radio wavelengths with the Multi-Element Radio-Linked Interferometer Network (MERLIN), the Very Large Array (VLA), the Very Long Baseline Array (VLBA) and the European VLBI Network (EVN)[14–16].

Here we concentrate on the early Very Long Baseline Interferometry (VLBI) and MERLIN imaging observations that resolve the expanding radio source (Fig. 1). In the first epoch (13.8 d after outburst, taking day 0 as 2006 February 12.83; ref. 17) the radio emission takes the form of an approximately circular structure, which is significantly brighter on its eastern side. This one-sided ring initially expands at a speed of about 0.62 mas d$^{-1}$ (Fig. 2), equivalent to 1,730 km s$^{-1}$ in the plane of the sky at our assumed distance of 1,600 pc. However, the structure quickly grows more complex with the appearance of a second component to the east of the ring. Subsequent 5-cm MERLIN imaging also shows the appearance of a third component to the west (Fig. 3). In 1985, RS Oph was followed at radio wavelengths from 18 d to a year after outburst[18,19]. Only one attempt was made (77 d after outburst) to obtain a VLBI image[5,6], interpreted as a three-component radio source extending east–west to ~200 mas. As only three telescopes of the early EVN had been used, it was unclear whether this reflected the true nature of the emission. However, our results suggest that the source is developing into a similar structure during the current outburst.

Our measurement of the expansion speed of the ring is consistent with velocity estimates from the Doppler-broadened line profiles seen at optical[20], infrared[21] and X-ray[10] wavelengths and in ultraviolet spectroscopy[22] of the 1985 outburst. Assuming this corresponds to the velocity of the shock wave resulting from the explosion, then gas would be heated to ~$4\times10^7$ K, consistent with temperatures in the range $(0.2–7)\times10^7$ K inferred from X-ray observations with the Swift, Chandra and RXTE satellites[10,13,23]. This scenario is also consistent with our observed brightness temperatures. Taking the ratio of mass-loss rate (into the red giant wind) to wind velocity to be $6\times10^{12}$ g cm$^{-1}$ (from models of the 1985 outburst[4]), we estimate the density in the shell 14 d after outburst to be around $10^{-17}$ g cm$^{-3}$. A shell with these densities and a temperature of around $10^7–10^8$ K has a very small optical depth to free–free radiation. In such a case, the observed brightness temperatures are much lower than the electron temperature, and so thermal emission from hot gas alone is insufficient to explain the observed level of radiation. We therefore conclude that there is a significant contribution from a non-thermal mechanism, probably synchrotron radiation from particles accelerated in the shock wave in a manner similar to that inferred for supernova remnants.

The one-sided appearance of the ring is likely to be due to the density distribution in the red giant wind and/or asymmetries in the explosion. Notably, the second component is visible at first only in the 6-cm image but a week later is also visible at 18 cm, indicating its spectrum has flattened, consistent with a source that is initially subject to free–free absorption

(greater at longer wavelengths) in some overlying material. The likely source of such absorption is the ionized red giant wind. Figure 4 presents a simple model, where the nova explosion results in a bipolar shock-heated shell that expands through the red giant wind. The bipolarity is the result of either an inherent asymmetry in the mass ejection from the white dwarf (that is, a jet-like outflow), or an equatorial enhancement in the red giant wind that leads to shaping of the shock wave in a manner similar to the interacting stellar winds model for planetary nebulae[24]. Whatever the origin of the bipolarity, it is reasonable to expect its symmetry axis to be perpendicular to the plane of the orbit of the central binary system. The inclination of the binary orbit to the line of sight is estimated to be 30–40° from radial velocity observations before the 2006 outburst[25], so we adopt a value of 35°. Figure 4 shows model images at around the time of the VLBI imaging and the later MERLIN imaging. The early one-sided structure is quite clear, and the model predicts that as the source expands, the free–free absorption reduces, so the structure becomes more symmetrical with a central bright source between two other components, just as the MERLIN imaging (Fig. 3) is suggesting and as was seen at day 77 in 1985[5,6].

Although the binary inclination will remain the same from one outburst to the next, the observed structure may also depend on the relative orientation of the binary components at the time of outburst. Orbital elements determined from infrared spectroscopy[26] suggest that the 2006 outburst took place when the red giant and white dwarf appear most widely separated. With an orbital period of 456 d this is about the same phase as the 1985 outburst 16.9 orbits earlier. Hence we expect the geometrical properties to be comparable, consistent with the similar radio structures seen in the two outbursts.

The model also explains the rapid rise in the radio emission—detected by MERLIN only 4.4 d after outburst[14], and projecting back to a turn-on at day 3.8—as the result of rapidly reducing absorption in the central, densest, regions of the red giant wind. However, a uniformly expanding shell is not consistent with the slow movement of the second component. More detailed analysis, including a physically self-consistent description of the density and temperature distribution within the shell and recognition that the explosion on the white dwarf does not lie at the centre of the red giant wind, requires a full hydrodynamical simulation, which we leave for a paper to follow.

These observations provide an elegant confirmation of the basic shock model for RS Ophiuchi developed after the 1985 outburst, but provide new challenges in understanding the unexpected asymmetries seen in the radio shell. Their development into a two-sided structure is the best evidence thus far for the existence of jets in a recurrent nova. The mechanisms needed to produce jets in a wide variety of sources (ranging from active galactic nuclei to young stellar objects) are still unclear, but the favoured model suggests production at the centre of accretion disks with hydromagnetic acceleration and collimation[27]. This is consistent with our model, in which the symmetry axis of the bipolar shell is orthogonal to the plane of the binary orbit.

Further epochs of high-resolution radio imaging, combined with ongoing X–ray monitoring, will enable the shock dynamics in a nova explosion to be studied in unprecedented detail, including modelling the departure from spherical symmetry. As well as providing additional insights into important physical processes such as particle acceleration, this will lead to improved estimates of the ejected mass and outburst energy, essential to constraining the thermonuclear runaway model and to determining whether RS Oph and objects of its type will eventually explode as type Ia supernovae[28].

**Acknowledgements** We are grateful to the staff of the VLBA (a facility of the NSF operated under cooperative agreement by Associated Universities, Inc.), EVN (a joint facility of European, Chinese, South African and other radio astronomy institutes funded by their national research councils), the NASA DSN facility at Robledo and MERLIN (a National Facility of PPARC operated by the University of Manchester) for help in obtaining these observations at short notice.

**Author Information** Reprints and permissions information is available at npg.nature.com/reprintsandpermissions. The authors declare no competing financial interests. Correspondence and requests for materials should be addressed to T.J.O'B. (tim.obrien@manchester.ac.uk).

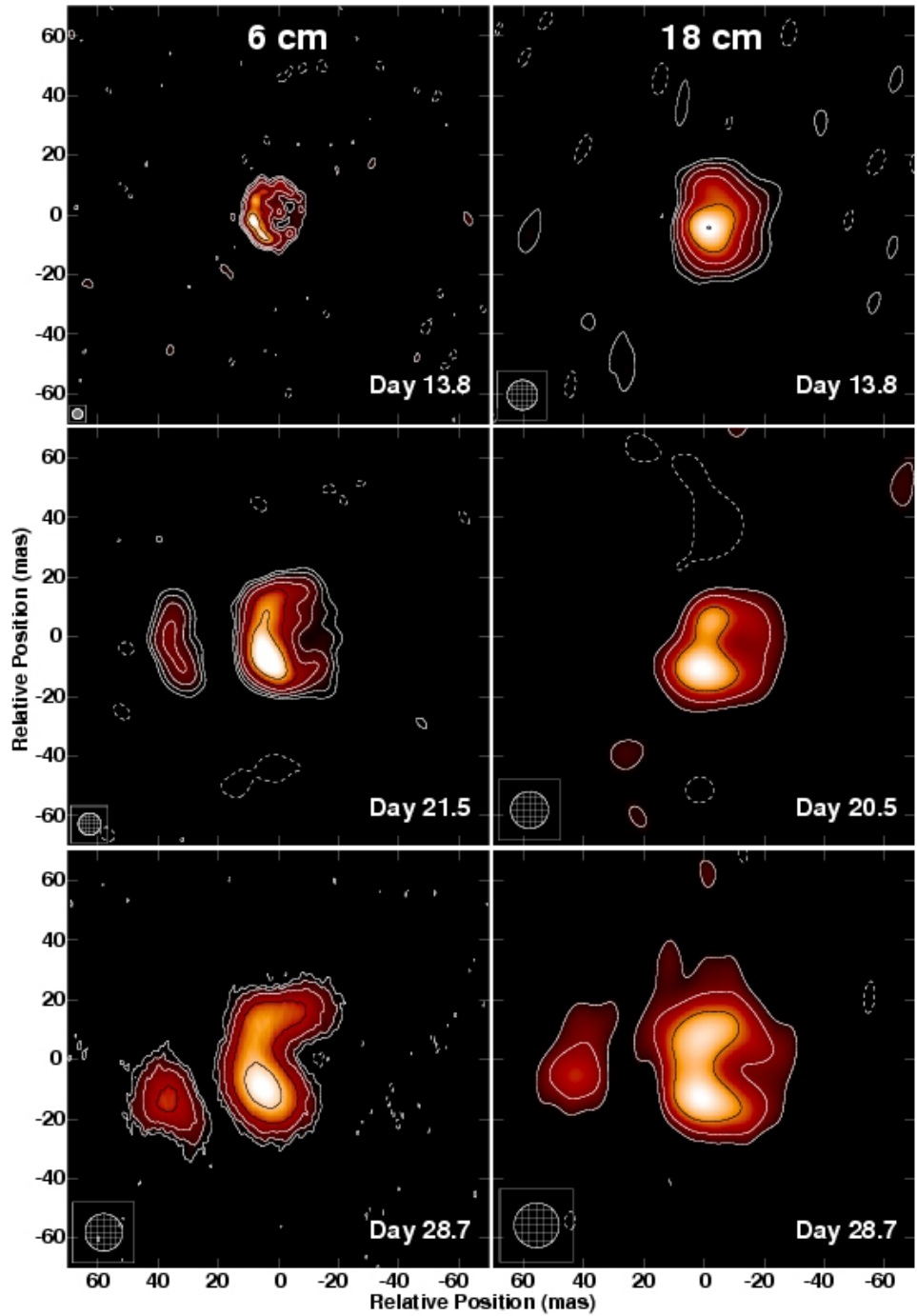

**Figure 1 High-resolution radio images of RS Oph.** Images of RS Oph at wavelengths of 6 cm (left column) and 18 cm (right column) made with the VLBA on 2006 February 26 (day 13.8) and March 13 (day 28.7), and the EVN on March 5/6 (day 20.5/21.5). In each case, north is up and east is to the left, and the images are restored with the circular beams shown at lower left. The first 6-cm VLBA image has a resolution of 3.3 mas (5 AU at a distance of 1,600 pc) and a peak flux density of 4.7 mJy beam$^{-1}$ corresponding to a brightness temperature of around $4\times10^7$ K. The contour levels are $(-1, 1, 2, 4, 8, 16, 32)$ times a base level given by 227, 105 and 220 µJy beam$^{-1}$ for the 6-cm images (in increasing time order), and 500, 900 and 900 µJy beam$^{-1}$ for the 18-cm images.

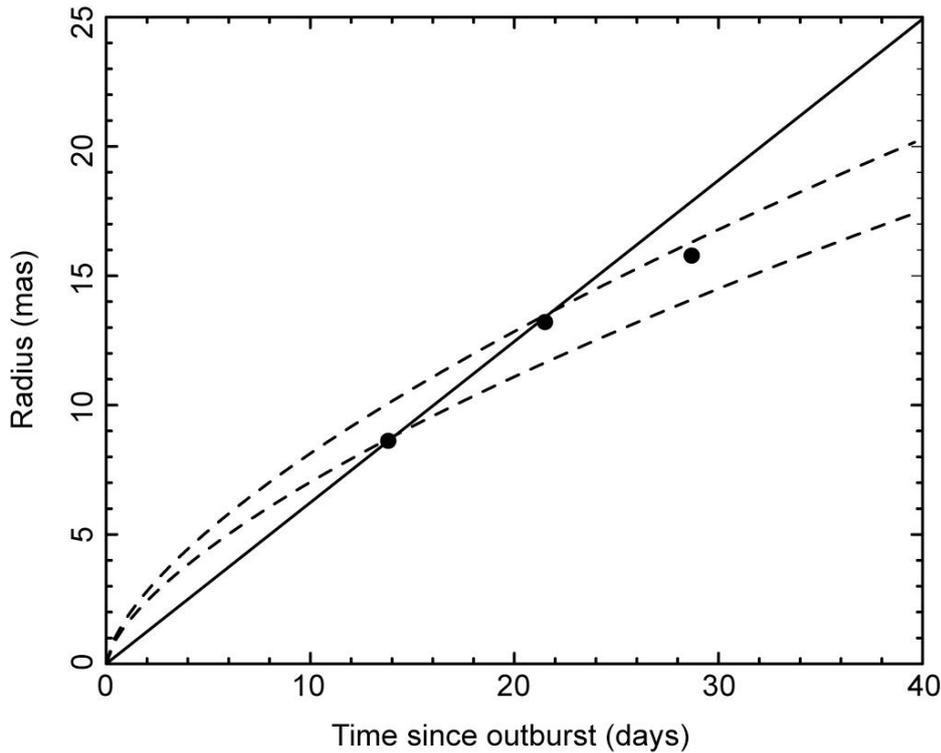

**Figure 2 Expansion of the bright ring-like component.** The plotted points are estimates of the radius of the ring-like component from the 6-cm images. The solid line drawn through the origin and the first data point corresponds to free expansion at a speed of 0.62 mas d$^{-1}$, equivalent to 1,730 km s$^{-1}$ at our assumed distance of 1,600 pc (ref. 19). However, existing models[8] suggest that after about day 6 the shock should be decelerating as it sweeps up red giant wind with its radius given by $at^{2/3}$ (here $a$ is a constant, and $t$ is time since outburst). Comparison with the X-ray emission seen in 1985[4] led to estimates for $a$ of $2.03 \times 10^{10}$ cm s$^{-2/3}$ and $2.35 \times 10^{10}$ cm s$^{-2/3}$, shown as the dashed lines. These are broadly consistent with the observations. However, although the expansion is clear, the measurements are sufficiently difficult that at this stage we are not able to clearly distinguish free expansion from the predicted deceleration. The 6-cm image from 26 February is clearly ring-like and a radius can easily be measured, but in the following two epochs the analysis is harder as the ring is only one-sided. Here we adopt the following method. In each 6-cm image, a series of north–south slices show two peaks, one from each side of the ring. The plotted points are half the separation of these peaks, determined by fitting gaussian profiles to the slice showing the largest separation. Formal errors on these measurements are typically around 0.2 mas, but these underestimate the true uncertainty. For example, analysis of the expanding shell of supernova 1993J[29] shows that the peaks may not represent the true location of the shock wave as they will be a convolution of the underlying emission with the beam.

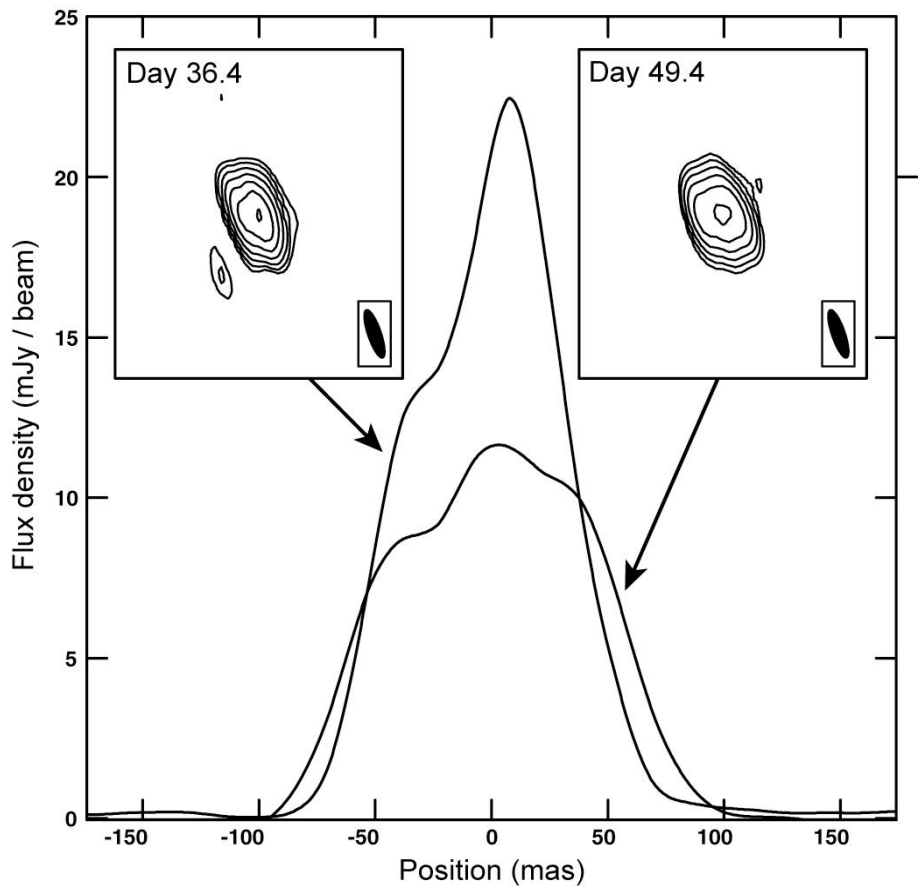

**Figure 3 A third component seen in MERLIN radio imaging.** East–west slices through the MERLIN maps (at a wavelength of 5 cm, shown as insets) of RS Oph from 21 March (day 36.4) and 3 April (day 49.4). These show that on day 36.4 the source comprised a bright component (which we identify with the main ring in the images of Fig. 1) and a fainter component to the east (the feature which first appears in the EVN map of day 21.5). By day 49.4 the slice shows that MERLIN has detected a third component to the west. The contours in the maps are at (−1, 1, 2, 4, 8, 16, 32, 64)×326 µJy beam$^{-1}$, and the beam size is 121×31 mas.

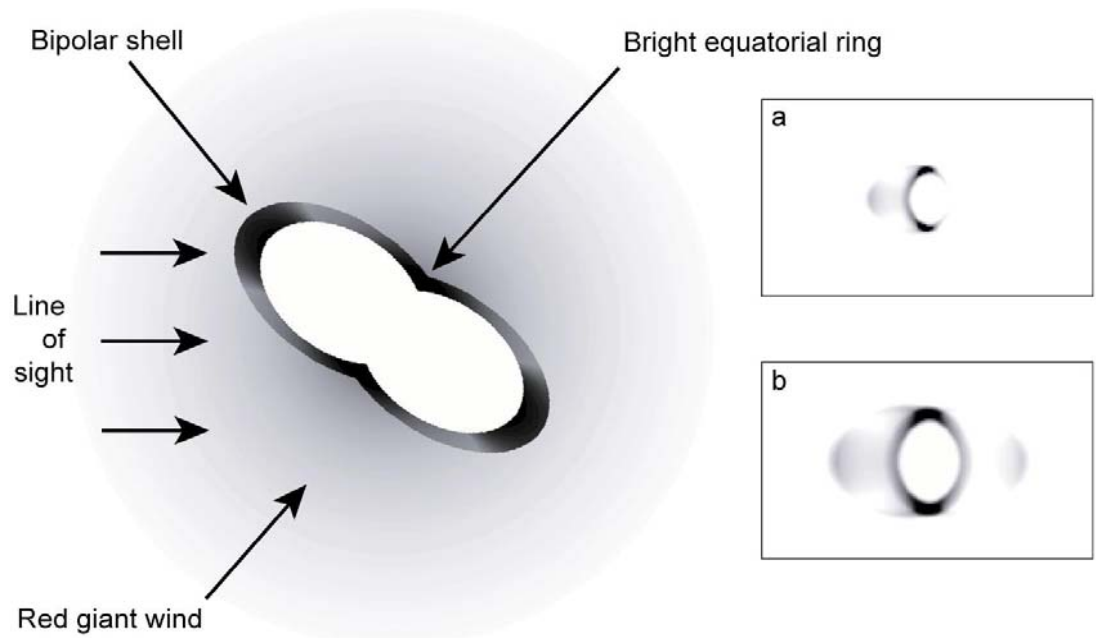

**Figure 4 A bipolar model for the radio emission.** Main figure, a slice showing gas density in a simple model for the structure of RS Oph. A bipolar shell of shock-heated gas, swept up by the ejecta, is embedded within a sphere filled with 21 yr of mass-loss from the red giant. The gas density in the red giant wind, given by the 1985 model fits[4], is proportional to $r^{-2}$ (where $r$ is radius) and its temperature is $10^4$ K. The density in the shocked shell is simply assumed to be four times the wind density at that radius, but with some enhancement towards the poles, and its temperature is $5\times10^7$ K. The gas emits via a combination of free–free and flat-spectrum synchrotron emission in which we assume equipartition between the energy in relativistic electrons (a fixed small fraction of the thermal energy) and the magnetic field. The emission is integrated along a line of sight given by the inclination of the binary system allowing for intrinsic free–free absorption, mostly provided by the ionized red giant wind. Synthetic images are shown as insets, oriented for comparison with the observations. The central ring is the waist of the shell, which is the densest, and hence brightest, part. In the first epoch (top inset), the furthest half of the ring actually suffers less absorption owing to the swept-out cavity along this line of sight, and so appears brighter. The second component, first seen at day 21.5, is the nearer lobe of the shell with the farther lobe initially obscured. At the second epoch (bottom inset) the shell has expanded, the absorption is reduced and the appearance has become more symmetrical. We identify the third component seen in the second MERLIN map of Fig. 3 with the unveiling of the more distant lobe.